# Performance of Cooperative NOMA with MRT/RAS in Nakagami-m Fading Channels with Channel Estimation Errors

Mahmoud Aldababsa, Oğuz Kucur

*Abstract*— This paper studies the performance of a downlink non-orthogonal multiple access (NOMA) based cooperative network with maximal ratio transmission/receive antenna selection (MRT/RAS) over Nakagami-*m* fading channels in the presence of channel estimation errors (CEEs). In the system, a base station communicates with multiple mobile users through a half duplex channel state information based amplify-and-forward relay. All nodes are equipped with multiple antennas and the hybrid diversity technique MRT/RAS is employed in both hops. The outage behavior of the system is investigated by driving closed-form expression for outage probability (OP). In addition, the corresponding lower and upper bounds of the derived OP are obtained. Moreover, the behavior of the system is studied in high signal-to-noise ratio region by obtaining an error floor value in the presence of CEE as well as achieving diversity and array gains in the absence of CEE. Finally, the analytical results in the presence and absence of the CEEs are verified by the Monte Carlo simulations. Results show that the MRT/RAS scheme enhances the OP significantly and is much more robust to the CEEs in comparison with the single antenna case.

## I. INTRODUCTION

With the rapidly increase in the development of Internet mobile data and Internet of things (IoT), more devices and/or users and much higher achievable data rates pose huge challenges for the future wireless mobile communication systems such as fifth generation (5G). The 5G and beyond aim to support quite higher spectral efficiency to handle the explosive data traffic and more massive connectivity of the user equipments to fulfill ultra low latency. In order to satisfy the aforementioned challenges, non-orthogonal multiple access (NOMA) is strongly considered as a nominee and eligible multiple access technology [1]-[2]. It can remarkably enhance overall spectral efficiency of the system and quality of service (QoS). It is somehow different from the conventional orthogonal multiple access (OMA) in which every user can use only a specific partition of available orthogonal resources, i.e., frequency/time/code and also the receiver of each user takes the advantage of the orthogonality in order to separate its own signal out from others. However, the key feature of the NOMA is to serve multiple users over the non-orthogonal resources [3]. In particular, the power domain downlink NOMA superposes all the users' information signals at the transmitter while at the receivers the successive interference cancellation process is successively curried out to reduce the inter-user interference and recover the desired users' signals [4].

Multiple-input multiple-output (MIMO) technology can play a crucial role for improving the whole wireless performance of the system in the terms of capacity and error probability [5]. Especially, antenna diversity techniques have been implemented at transmitter and/or receiver sides in different ways diminishing the ruinous effect of the wireless communication channels [6]. Due to these attractive advantages, the MIMO techniques have been integrated into the NOMA systems in [7]-[14]. In [7], a transmit antenna selection (TAS) scheme is proposed in the downlink NOMA network. Its selection criterion depends on choosing the best transmit antenna providing the largest sum rate and highest Jain's fairness index. Different from [7], to optimize the performance of the security of the NOMA network, the TAS is applied in [8]-[9] by selecting the best antenna offering higher security. Moreover, new joint antenna selection schemes, namely max-max-max and max-min-max have been designed in [10]-[12], where the sum rate performance is investigated for both schemes. However, the max-max-max scheme intends to enhance the good user's performance while the max-min-max is designed to improve the bad user's performance. Furthermore, in [13], maximum ratio transmission (MRT) is applied to the NOMA network to have maximum sum rate. Also, another transmit diversity technique, conventional Alamouti space-time block coding (STBC) has been applied to the downlink NOMA in [14].

On the other hand, cooperative relaying is also a substantial and beneficial technique in wireless systems because of its large coverage area without needing extra transmission power in addition to the improvement of the error performance [15]. Due to these considerable advantages, cooperative relaying has been applied to the NOMA. However, the basic structure of the cooperative relaying is a dual-hop relaying network where the base station (BS) communicates with the users by the help of a relay. There are no direct links between the BS and users, and hence the total transmission process from the BS to the users is split into two time slots. In the first time slot, the BS transmits the message signal to the relay; in the second time slot the relay forwards the message signal to the users. In the decode-and-forward (DF) relaying the relay decodes the received signal, and then forwards it to the users whereas the relay amplifies the message signal before forwarding to the users for amplify-and-forward (AF) relaying. The signal amplification in the AF relaying can be done in two ways, namely fixed [16] and channel state information (CSI) based [17] gains. In the fixed gain case, it is a constant expected value of the fading channel gains. On the other side, the amplification in the CSI based relaying varies according to the fading channel gains. In this context, the NOMA networks with relaying have been studied in [18]-[23]. Particularly, the authors have studied the performance of the cooperative NOMA in terms of outage probability (OP) and/or achievable data rate with the CSI based AF relaying in [18], the fixed gain relaying in [19] and different relay selection schemes in [20]-[23]. Meanwhile, performance of the cooperative MIMO-NOMA has also been studied in [24]-[28]. The authors examined the performance of the cooperative NOMA in [24]-[25], where the TAS is employed at the BS and maximal ratio combining (MRC) at mobile users. In addition, the performance of the Alamouti STBC scheme has been investigated for a single user NOMA with cooperative DF network in [26] and NOMA with cooperative spectrum sharing protocol in [27].

Moreover, in [28], the performance of full-duplex NOMA combined with MRT or MRC is studied in terms of the sum rate. To furthermore minimize the OP of multiple-relay NOMA network, joint relay-and-antenna selection scheme is proposed in [29]-[31]. However, in [29], a multi-relay DF cooperation network is considered with multi-antenna BS and two mobile users. The optimal relay, the best transmit antenna at the source, and the optimal receive antennas at the mobile users are jointly determined. On the other hand, in [30], a multi-relay MIMO-NOMA network is also investigated. However, the TAS/receive antenna selection (RAS) is applied in the 1st hop between the BS and relays by selecting the best transmit-receive antenna pairs. While, the TAS/MRC scheme is employed in the 2nd hop by choosing the best transmit antennas of the relays for the links between the relays and strong user. Then, the relay is selected offering the maximum end-to-end signal-to-noise ratio (SNR). In [31], the joint relay-and-antenna selection is also proposed for AF cognitive radio inspired MIMO-NOMA network satisfying higher QoS besides better outage performance. Despite most of NOMA systems have been studied in ideal cases, practical cases such as channel estimation errors (CEEs) have also been investigated in the cooperative NOMA in [32] and cooperative MIMO-NOMA in [33], in which the TAS and MRC are applied at the BS and mobile users, respectively.

The performance of NOMA with the MRT diversity technique, which is a special case of the beamforming, has been studied in a few works, especially in [13] and [28] in the ideal case. The MRT provides optimum performance but with requiring full CSI causes much more complexity at the transmitter, therefore it is convenient for simple receivers. However, hybrid diversity scheme MRT/RAS, which is a combination of the MRT at the transmitter and RAS at the receiver, can be a considerable benefit for wireless communication system designs since it provides good performance and tolerates some hardware complexity at the receiver due to the MRT. In addition, the RAS offers simplicity at the receiver as it utilizes just one radio frequency (RF) chain for antenna providing the maximum channel gain. To the best of our knowledge from the literature, the performance of the hybrid diversity scheme MRT/RAS has not been investigated in cooperative NOMA networks even in the absence of the CEE. With this motivation, we study the performance of the downlink cooperative NOMA network with MRT/RAS in the presence of the CEE. In the network, the communication between BS with multi-transmit antenna and mobile users with multi-receive antenna is done through CSI based AF relay with multi-receive and multi-transmit antenna. The MRT/RAS scheme is employed in both hops. The key contributions of this paper can be highlighted as follows: 1) The outage performance of the network is provided in the presence and absence of the CEE over Nakagami-$m$ fading channels. 2) The closed-form expression of the OP has been derived. 3) The lower and upper bounds of the OP are obtained. 4) In case of the imperfect CSI, the OP is evaluated at the high signal-to-noise ratio (SNR) region in order to observe the presence of an error floor (EF). 5) In case of the perfect CSI, an asymptotic analysis is carried out such that the diversity and array gains are obtained. 6) The theoretical results are validated by simulations and demonstrate superiority of the proposed NOMA with MRT/RAS over the conventional OMA with MRT/RAS.

The remainder of this paper is organized as follows: In Section II, the system and channel models are described. In Section III, the performance analysis is conducted. The numerical results are presented in Section IV to confirm the analysis, and the paper is concluded in Section V.

**Notations:** Throughout this paper, $E[.]$ denotes the expectation operator; $\|.\|$ denotes Frobenius norm; $I_N$ refers the identity matrix of order $N$; $f_X(.)$ and $F_X(.)$ denote the probability density function (PDF) and cumulative distribution function (CDF) of a random variable $X$, respectively; $\Psi(.,.)$ and $\Gamma(.)$ are lower incomplete and Gamma functions, respectively; $K_q(.)$ is modified Bessel function with $q$th order and second kind; $P_r(.)$ symbolizes probability and $(.)^H$ represents the Hermitian transpose. System and channel models

## II. SYSTEM AND CHANNEL MODELS

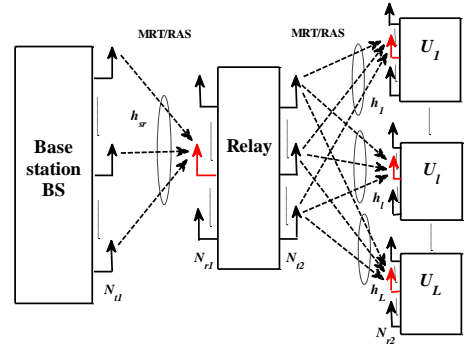

Figure 1: System model.

We consider the system model given in Fig. 1. The BS and each of the $L$ mobile users are equipped with $N_{t1}$ transmit and $N_{r2}$ receive antennas, respectively, and the half duplex CSI based AF relay is equipped with $N_{r1}$ receive and $N_{t2}$ transmit antennas. Since there are large distances leading to a large path loss effect or blockages by the large obstacles, deep fading and/or high shadowing effects between BS and users, the direct links are neglected. Also, a homogeneous network topology is considered, where all the mobile users are clustered relatively close together. The hybrid MRT/RAS scheme is employed in both hops such that single receiver antennas providing the highest total channel gains are selected for the relay in the 1st hop and each mobile user in the 2nd hop independently. The BS is assumed to have knowledge of necessary CSI and order of channel quality for the links between the relay and mobile users. During the training period, by assuming that uplink and downlink channels are reciprocal, in the 2nd hop each mobile user forwards pilot symbols to the relay, whereas in the 1st hop the BS forwards pilot symbols to the relay. Then, the relay estimates the channel gains of the 2nd hop, and determines and forwards indices of the best receive antennas of the 2nd hop to the mobile users. The relay also forwards the code of a look-up table that represents the order of the channel gains of the 2nd hop and the estimated coefficients of the 1st hop to the BS. We also assume ideal feedback channels.

Define $\boldsymbol{h}_{sr}$ as the fading channel coefficient vector ($N_{t1} \times 1$) in the 1st hop, which is between the BS and selected receive antenna of the relay, $\hat{\boldsymbol{h}}_{sr}$ is the ($N_{t1} \times 1$) estimated channel coefficient vector of $\boldsymbol{h}_{sr}$, $\boldsymbol{e}_{sr}$ is the ($N_{t1} \times 1$) CEE vector which is statistically independent of $\hat{\boldsymbol{h}}_{sr}$ and modelled as a complex Gaussian random variable [34] with zero mean and $E[\boldsymbol{e}_{sr}\boldsymbol{e}_{sr}^H] = \boldsymbol{I}_{N_{t1}}\sigma_{e,sr}^2$, where $\sigma_{e,sr}^2$ is the variance of each element in $\boldsymbol{e}_{sr}$. Similarly, the ($N_{t2} \times 1$) fading channel coefficient vector in the 2nd hop is $\boldsymbol{h}_l$, which is between the relay and selected receive antenna of the $l$th user, $\hat{\boldsymbol{h}}_l$ refers to the ($N_{t2} \times 1$) estimated channel coefficient vector of $\boldsymbol{h}_l$, $\boldsymbol{e}_l$ is the ($N_{t2} \times 1$) CEE vector which is statistically independent of $\hat{\boldsymbol{h}}_l$ and modelled as complex Gaussian noise with zero mean and $E[\boldsymbol{e}_l\boldsymbol{e}_l^H] = \boldsymbol{I}_{N_{t2}}\sigma_{e,l}^2$, where $\sigma_{e,l}^2$ is the variance of each element in $\boldsymbol{e}_l$. The optimal transmit weight vectors are expressed as $\boldsymbol{w}_{sr} = \frac{\boldsymbol{h}_{sr}^H}{\|\boldsymbol{h}_{sr}\|}$ and $\boldsymbol{w}_l = \frac{\boldsymbol{h}_l^H}{\|\boldsymbol{h}_l\|}$, where $\|\boldsymbol{h}_{sr}\| = \sqrt{\sum_{i=1}^{N_{t1}}|h_{sr}^{(i,j_{sr})}|^2}$, $h_{sr}^{(i,j_{sr})}$ is the channel gain between the $i$th transmit antenna of the BS and the selected receive antenna ($j_{sr}$th) of the relay, $\|\boldsymbol{h}_l\| = \sqrt{\sum_{i=1}^{N_{t2}}|h_l^{(i,j_l)}|^2}$ and $h_l^{(i,j_l)}$ is the channel gain between the $i$th transmit antenna of the relay and the selected receive antenna ($j_l$th) of the $l$th mobile user. Now, the whole communication process between the BS and mobile users is completed in two time slots. In the first one, the BS transmits the superimposed signal $\boldsymbol{w}_{sr}\sum_{i=1}^{L}\sqrt{P_s a_i}x_i$ to the relay. Here, $x_i$ denotes the $i$th user' information ($U_i$) of unit energy, $P_s$ refers the transmit power at the BS and $a_i$ is the $i$th user' power allocation factor such that $a_1 + \ldots + a_L = 1$ and $a_1 \geq a_2 \geq \ldots \geq a_L$ opposite to the order of the channel gains of the 2nd hop due to the NOMA concept. Then, at the selected receive antenna of the relay, the received signal can be expressed as follows:

$$y_r = \boldsymbol{h}_{sr}\boldsymbol{w}_{sr}\sum_{i=1}^{L}\sqrt{P_s a_i}x_i + n_r$$
$$= \|\boldsymbol{h}_{sr}\|\sum_{i=1}^{L}\sqrt{P_s a_i}x_i + n_r$$
$$= \|\hat{\boldsymbol{h}}_{sr} + \boldsymbol{e}_{sr}\|\sum_{i=1}^{L}\sqrt{P_s a_i}x_i + n_r, \quad (1)$$

where $n_r$ is zero mean complex additive Gaussian noise with a variance of $\sigma_r^2$ at the selected receive antenna of the relay. In the 2nd hop, the relay amplifies the weighted signal $\boldsymbol{w}y_r = \left(\frac{1}{L}\sum_{j=1}^{L}\boldsymbol{w}_j\right)y_r$ with an amplification factor $G = \frac{1}{\sqrt{P_s\|\hat{\boldsymbol{h}}_{sr}\|^2 + P_s\sigma_{e,sr}^2 + \sigma_r^2}}$. Thus, at the selected receive antenna of the $l$th user, the received signal can be expressed as

$$y_l = \boldsymbol{h}_l G\sqrt{P_r}\boldsymbol{w}y_r + n_l$$

$$= G(\hat{\boldsymbol{h}}_l + \boldsymbol{e}_l) \|\hat{\boldsymbol{h}}_{sr} + \boldsymbol{e}_{sr}\| \boldsymbol{w} \sum_{i=1}^{L}\sqrt{P_s P_r a_i}x_i$$
$$+ G\sqrt{P_r}(\hat{\boldsymbol{h}}_l + \boldsymbol{e}_l)\boldsymbol{w}n_r + n_l, \quad (2)$$

where $P_r$ denotes the transmission power at the relay and $n_l$ is the zero mean complex additive Gaussian noise with a variance of $\sigma_l^2$. In (??), without loss of generality, the channel gains are ordered as $\|\boldsymbol{h}_1\|^2 \leq \|\boldsymbol{h}_2\|^2 \leq \ldots \leq \|\boldsymbol{h}_L\|^2$. In the 1st and 2nd hops, the estimated channel coefficients between the $i$th antenna of the BS and $j$th receive antenna of the relay, and between the $i$th transmit antenna of the relay and $j$th antenna of the $l$th user are denoted by $\hat{h}_{sr}^{(i,j)}$ and $\hat{h}_l^{(i,j)}$, respectively. Magnitudes of the fading gains are identically and independently distributed (i.i.d.) with Nakagami-$m$ distribution. With the benefit of series expansion of Gamma [5, eq.(8.310.1)], incomplete Gamma [5, eq.(8.352.6)] and lower incomplete Gamma [5, eq.(8.350.1)] functions, the PDF and CDF of Gamma random variable $\hat{X}$, which is the square of Nakagami-$m$ random variable can be defined as

$$f_{\hat{X}}(x) = \left(\frac{m_x}{\hat{\Omega}_x}\right)^{m_x} \frac{x^{m_x-1}}{\Gamma(m_x)}e^{-\frac{m_x x}{\hat{\Omega}_x}} \text{ and } F_{\hat{X}}(x) = \frac{\Psi\left(m_x,\frac{m_x x}{\hat{\Omega}_x}\right)}{\Gamma(m_x)} = 1 - e^{-\frac{m_x x}{\hat{\Omega}_x}}\sum_{k=0}^{m_x-1}\left(\frac{m_x x}{\hat{\Omega}_x}\right)^k\frac{1}{k!},$$ respectively. Where, $m_x$ is the

parameter of Nakagami-$m$ distribution and $\hat{\Omega}_x = E[|\hat{X}|^2]$. Since the hybrid MRT/RAS scheme is employed in both hops, only single receiver antennas offering the maximum total of squared channel gains are selected for the relay in the 1st hop and each mobile user in the 2nd hop separately. According to this selection criterion, the index of the selected receive antenna at the relay in the 1st hop $j_{sr}$ and at mobile user $l$ in the 2nd hop $j_l$ can be stated, respectively as

$$j_{sr} = \arg\max_{1 \leq j \leq N_{r1}}\{\sum_{i=1}^{N_{t1}}|\hat{h}_{sr}^{(i,j)}|^2\}, \quad (3)$$

$$j_l = \arg\max_{1 \leq j \leq N_{r2}}\{\sum_{i=1}^{N_{t2}}|\hat{h}_l^{(i,j)}|^2\}, l = 1,2,\ldots,L. \quad (4)$$

The CDF of the squared channel gain in the 1st hop $\varphi_{sr} = \|\hat{\boldsymbol{h}}_{sr}\|^2$ can be given, according to the highest order statistics [36] as

$$F_{\varphi_{sr}}(x) = \left[1 - e^{-\frac{m_1 x}{\hat{\Omega}_1}}\sum_{k=0}^{m_1 N_{t1}-1}\left(\frac{m_1 x}{\hat{\Omega}_1}\right)^k\frac{1}{k!}\right]^{N_{r1}}, \quad (5)$$

where $m_1$ is the Nakagami channel parameter of the 1st hop, $\hat{\Omega}_1 = E[\|\hat{\boldsymbol{h}}_{sr}\|^2]$. Assumed $\alpha$ and $d_1$ denote path loss exponent and normalized distance between the BS and relay, receptively. Then, $\Omega_1 = E[\|\boldsymbol{h}_{sr}\|^2] = d_1^{-\alpha}$, and the relative CEE for the 1st hop is $\epsilon_{sr} = \frac{\sigma_{e,sr}^2}{\Omega_1}$. Hence, $\hat{\Omega}_1 = (1 - \epsilon_{sr})d_1^{-\alpha}$. Moreover, the PDF and CDF of the unordered random variable in the 2nd hop $\tilde{\varphi}_l = \|\hat{\boldsymbol{h}}_l\|^2$ can be expressed, respectively as

$$f_{\tilde{\varphi}_l}(x) = N_{r2}\left(\frac{m_2}{\hat{\Omega}_2}\right)^{m_2 N_{t2}}\frac{x^{m_2 N_{t2}-1}}{\Gamma(m_2 N_{t2})}e^{-\frac{m_2 x}{\hat{\Omega}_2}}$$
$$\times \left[1 - e^{-\frac{m_2 x}{\hat{\Omega}_2}}\sum_{k=0}^{m_2 N_{t2}-1}\left(\frac{m_2 x}{\hat{\Omega}_2}\right)^k\frac{1}{k!}\right]^{N_{r2}-1}, \quad (6)$$

$$F_{\tilde{\varphi}_l}(x) = \left[1 - e^{-\frac{m_2 x}{\hat{\Omega}_2}} \sum_{k=0}^{m_2 N_{t2}-1} \left(\frac{m_2 x}{\hat{\Omega}_2}\right)^k \frac{1}{k!}\right]^{N_{r2}}, \quad (7)$$

where $m_2$ is the Nakagami channel parameter of the 2nd hop, $\hat{\Omega}_2 = E[\|\hat{h}_l\|^2]$. Assume that $\Omega_2 = d_2^{-\alpha}$, where $d_2$ represents the normalized distance between the relay and mobile users. Since the relative CEEs for the 2nd hop are $\epsilon_l = \frac{\sigma_{e,l}}{\Omega_2}$, hence $\hat{\Omega}_2 = (1 - \epsilon_l) d_2^{-\alpha}$. If $Q_l = \frac{L!}{(L-l)!(l-1)!}$, then the PDF and CDF of ordered random variable $\varphi_l = \|\hat{h}_l\|^2$ can be given, respectively as [37]

$$f_{\varphi_l}(x) = Q_l \sum_{t=0}^{L-l} (-1)^t \binom{L-l}{t} f_{\tilde{\varphi}_l}(x) [F_{\tilde{\varphi}_l}(x)]^{l+t-1}. \quad (8)$$

$$F_{\varphi_l}(x) = Q_l \sum_{t=0}^{L-l} \frac{(-1)^t}{l+t} \binom{L-l}{t} \times [F_{\tilde{\varphi}_l}(x)]^{l+t}, \quad (9)$$

### III. PERFORMANCE ANALYSIS

#### A. Outage Probability

In order to derive the expression of the OP, first the expressions of signal-to-interference-and-noise ratio (SINR) should be obtained. Hence, from (2), *the exact expression of the instantaneous SINR of the $U_l$ to decode the $U_j$ can be derived as*

$$\gamma_{j,l}^{ex} = \frac{a_j \gamma^2 \varphi_{sr} \varphi_l \tilde{W}}{\gamma^2 \varphi_{sr} \varphi_l \tilde{W} \Sigma_j + \gamma \varphi_{sr} \tilde{\alpha}_1 + \gamma \varphi_l \tilde{W} \tilde{\alpha}_2 + \tilde{\alpha}_3}, \quad (10)$$

where $\gamma = P/\sigma^2$ is the SNR, $\sigma_r^2 = \sigma_l^2 = \sigma^2$ is assumed for mathematical tractability, $P = P_s = P_r$, $\Sigma_j = \sum_{i=j+1}^{L} a_i$, $\tilde{\alpha}_1 = \gamma \sigma_{e,l}^2 \tilde{W} + 1$, $\tilde{\alpha}_2 = \gamma \sigma_{e,sr}^2 + 1$, $\tilde{\alpha}_3 = \gamma^2 \sigma_{e,l}^2 \sigma_{e,sr}^2 \tilde{W} + \gamma \sigma_{e,l}^2 \tilde{W} + \gamma \sigma_{e,sr}^2 + 1$ and $\tilde{W} = \frac{1}{L^2} \|\sum_{i=1}^{L} w_i\|^2$. Take advantage of triangle inequality $\tilde{W} \leq (\frac{1}{L} \sum_{i=1}^{L} \|w_i\|)^2 = 1$,

$$\gamma_{j,l}^{ex} \leq \frac{a_j \gamma^2 \varphi_{sr} \varphi_l}{\gamma^2 \varphi_{sr} \varphi_l \Sigma_j + \gamma \varphi_{sr} \alpha_1 + \gamma \varphi_l \alpha_2 + \alpha_3}, \quad (11)$$

where $\alpha_1 = \gamma \sigma_{e,l}^2 + 1$, $\alpha_2 = \gamma \sigma_{e,sr}^2 + 1$ and $\alpha_3 = \gamma^2 \sigma_{e,l}^2 \sigma_{e,sr}^2 + \gamma \sigma_{e,l}^2 + \gamma \sigma_{e,sr}^2 + 1$. Consequently, *the upper bound expression of the instantaneous SINR of the $U_l$ to decode the $U_j$, the $U_l$ to detect its own signal and upper bound SNR of the $U_L$, can be expressed, respectively as*

$$\gamma_{j,l}^{up} = \frac{a_j \gamma^2 \varphi_{sr} \varphi_l}{\gamma^2 \varphi_{sr} \varphi_l \Sigma_j + \gamma \varphi_{sr} \alpha_1 + \gamma \varphi_l \alpha_2 + \alpha_3}, j \neq L, j < l, \quad (12)$$

$$\gamma_l^{up} = \frac{a_l \gamma^2 \varphi_{sr} \varphi_l}{\gamma^2 \varphi_{sr} \varphi_l \Sigma_l + \gamma \varphi_{sr} \alpha_1 + \gamma \varphi_l \alpha_2 + \alpha_3}, l \neq L, \quad (13)$$

$$\gamma_L^{up} = \frac{a_L \gamma^2 \varphi_{sr} \varphi_l}{\gamma \varphi_{sr} \alpha_1 + \gamma \varphi_l \alpha_2 + \alpha_3}. \quad (14)$$

**Proposition 1:** *The approximated OP of the lth user in the presence of the CEEs is expressed in closed form as*

$$OP_l^{app} = 1 + \frac{2(m_2/\hat{\Omega}_2)^{m_2 N_{t2}}}{\Gamma(m_2 N_{t2})} Q_l \sum \Xi (-1)^{p+t+r} \vartheta_k(p, m_1 N_{t1}) \vartheta_s(r, m_2 N_{t2})$$
$$\times e^{-\frac{m_2(r+1)\mu_l \alpha_1}{\hat{\Omega}_2}} e^{-\frac{p m_1 \mu_l \alpha_2}{\hat{\Omega}_1}} \mu_l^{m_2 N_{t2}+s+k-k_1-k_2-1} \alpha_1^{m_2 N_{t2}+s-k_2-1}$$
$$\times \alpha_2^{k-k_1} (\mu_l (\gamma \mu_l \alpha_1 \alpha_2 + \alpha_3)/\gamma)^{k_1} (q_2/q_3)^{\frac{q_1}{2}} K_{q_1}(2\sqrt{q_2 q_3}). \quad (15)$$

In (15), $\sum = \sum_{p=1}^{N_{r1}} \sum_{k=0}^{p(m_1 N_{t1}-1)} \sum_{t=0}^{L-l} \sum_{r=0}^{N_{r2}(l+t)-1} \sum_{s=0}^{r(m_2 N_{t2}-1)} \sum_{k_1=0}^{k} \sum_{k_2=0}^{m_2 N_{t2}+s-1}$, $\Xi = \binom{N_{r1}}{p} \binom{L-l}{l} \binom{N_{r2}(l+t)-1}{r} \binom{k}{k_1} \binom{m_2 N_{t2}+s-1}{k_2}$, $\mu_l = \max_{1 \leq j \leq l} \{\frac{\gamma_{th_j}}{\gamma(a_j - \gamma_{th_j} \sum_{i=j+1}^{L} a_i)}\}$, $\gamma_{th_j}$ is the threshold value of the SNR for the $j$th user, $q_1 = k_2 - k_1 + 1$, $q_2 = \frac{m_2(r+1)}{\hat{\Omega}_2}$, $q_3 = \frac{p m_1 \mu_l (\gamma \mu_l \alpha_1 \alpha_2 + \alpha_3)}{\hat{\Omega}_1 \gamma}$ and $\vartheta_x(y, q_z)$ denotes multinomial coefficients. With $w_\beta = (q_z/\Omega_z)^\beta/\beta!$, $\vartheta_0(y, q_z) = 1$, and $\vartheta_x(y, q_z) = 0$ if $\beta > q_z - 1$ multinomial coefficients is given by [5, eq.(0.314)]

$$\vartheta_x(y, q_z) = \frac{1}{x w_0} \sum_{\beta=1}^{x} (\beta(y+1) - x) w_\beta \vartheta_{x-y}(y, q_z), x \geq 1. \quad (16)$$

And *the approximated OP of the lth user in the absence of the CEEs can be easily obtained by putting $\sigma_{e,sr}^2 = \sigma_{e,l}^2 = 0$, which results in $\alpha_1 = \alpha_2 = \alpha_3 = 1$ and then substituting $\alpha$s into (15).*

*Proof of Proposition 1:* In general, OP can be defined as

$$OP_l^{ex} = 1 - P_r(\gamma_{j,l}^{ex} \geq \gamma_{th_j}) \geq 1 - P_r(\gamma_{j,l}^{up} \geq \gamma_{th_j})$$
$$= P_r(\gamma_{j,l}^{up} < \gamma_{th_j}) \quad (17)$$

As will be seen in Section 4, the numerical results show that the lower bound OP is about the exact one, and hence it can be considered to be roughly approximated value to the exact one, i.e., $OP_l^{ex} \approx OP_l^{low} = OP_l^{app}$. By using (12), the approximated $OP^{app}$ of the $l$th user can be stated as

$$OP_l^{app} = P_r\left(\varphi_{sr} < \frac{\mu_l(\gamma \alpha_2 \varphi_l + \alpha_3)}{\gamma(\varphi_l - \mu_l \alpha_1)}, \varphi_l > \mu_l \alpha_1\right)$$
$$= \underbrace{\int_0^{\mu_l \alpha_1} f_{\varphi_l}(x) dx}_{I_1}$$
$$+ \underbrace{\int_{\mu_l \alpha_1}^{\infty} f_{\varphi_l}(x) F_{\varphi_{sr}}\left(\frac{\mu_l(\gamma \alpha_2 x + \alpha_3)}{\gamma(x - \mu_l \alpha_1)}\right) dx}_{I_2}. \quad (18)$$

By the help of the multinomial coefficients in (16) and using (5) and (8), $I_2$ in (18) can be restated as in (19), given at the top of next page.

$$I_2 = \underbrace{\int_{\mu_l \alpha_1}^{\infty} f_{\varphi_l}(x) dx}_{1-I_1}$$
$$+ \frac{N_{r2}(m_2/\hat{\Omega}_2)^{m_2 N_{t2}}}{\Gamma(m_2 N_{t2})} Q_l \sum_{t=0}^{L-l} \sum_{p=1}^{N_{r1}} \sum_{k=0}^{p(m_1 N_{t1}-1)} \sum_{t=0}^{L-l} \sum_{r=0}^{N_{r2}(l+t)-1} \sum_{s=0}^{r(m_2 N_{t2}-1)} (-1)^{t+r+p} \binom{N_{r1}}{p} \binom{L-l}{l} \binom{N_{r2}(l+t)-1}{r}$$
$$\vartheta_k(p, m_1 N_{t1}) \vartheta_s(r, m_2 N_{t2})$$
$$\times \underbrace{\int_{\mu_l \alpha_1}^{\infty} x^{m_2 N_{t2}+s-1} e^{-\frac{m_2(r+1)x}{\hat{\Omega}_2}} \left(\frac{\mu_l(\gamma \alpha_2 x + \alpha_3)}{\gamma(x - \mu_l \alpha_1)}\right)^k e^{-\frac{p m_1 \mu_l(\gamma \alpha_2 x + \alpha_3)}{\hat{\Omega}_1 \gamma(x - \mu_l \alpha_1)}} dx}_{I_3}$$

(19)

By substituting the variable $y = x - \mu_l \alpha_1$ in (19) and using the binomial theorem, then $I_3$ can be rewritten as

$$I_3 = e^{-\frac{m_2(r+1)\mu_l \alpha_1}{\hat{\Omega}_2}} e^{-\frac{p m_1 \mu_l \alpha_2}{\hat{\Omega}_1}} \sum_{k_1=0}^{k} \sum_{k_2=0}^{m_2 N_{t2}+s-1} \binom{k}{k_1}$$
$$\binom{m_2 N_{t2}+s-1}{k_2} (\mu_l \alpha_1)^{m_2 N_{t2}+s-k_2-1} y^{k_2} (\mu_l \alpha_2)^{k-k_1} \times$$
$$\left(\frac{\mu_l(\gamma \mu_l \alpha_1 \alpha_2 + \alpha_3)}{\gamma}\right)^{k_1} \underbrace{\int_0^{\infty} e^{-\frac{m_2(r+1)y}{\hat{\Omega}_2}} e^{-\frac{p m_1 \mu_l(\gamma \mu_l \alpha_1 \alpha_2 + \alpha_3)}{\hat{\Omega}_1 \gamma y}} y^{k_2-k_1} dy}_{I_4}.$$

(20)

$I_4$ in (20) can be expressed in the terms of the modified Bessel function $K_q(.)$ [5, eq.(3.471.9)]. Consequently, by substituting (19) and (20) into (18), the expression of the approximated OP in closed form is obtained as in (18) in the presence of the CEEs for the $l$th user.

### B. Bounds of Outage Probability

By reforming $OP_l^{app}$ in (18) and using $\frac{1}{2} \min(a, b) \leq \frac{ab}{a+b} \leq \min(a, b)$ [38], *the lower and upper bounds of the approximated OP for the lth user in the presence of the CEEs can be given, respectively by*

$$OP_{l,LB}^{app} = 1 - P_r \left( \min\left(\frac{\gamma \alpha_1}{\alpha_3} \varphi_{sr}, \frac{\gamma \alpha_2}{\alpha_3} \varphi_l\right) \geq \frac{\alpha_1 \alpha_2}{\alpha_3} \mu_l \right)$$
$$= F_{\varphi_{sr}}(\alpha_2 \mu_l) + F_{\varphi_l}(\alpha_1 \mu_l) F_{\varphi_{sr}}(\alpha_2 \mu_l) F_{\varphi_l}(\alpha_1 \mu_l), \quad (21)$$

$$OP_{l,UB}^{app} = 1 - P_r \left( \frac{1}{2} \min\left(\frac{\gamma \alpha_1}{\alpha_3} \varphi_{sr}, \frac{\gamma \alpha_2}{\alpha_3} \varphi_l\right) \geq \frac{\alpha_1 \alpha_2}{\alpha_3} \mu_l \right)$$
$$= F_{\varphi_{sr}}(2\alpha_2 \mu_l) + F_{\varphi_l}(2\alpha_1 \mu_l) F_{\varphi_{sr}}(2\alpha_2 \mu_l) F_{\varphi_l}(2\alpha_1 \mu_l). \quad (22)$$

By the help of the multinomial coefficients in (16) and substituting (5) and (9) into (21) and (22), we obtain the expressions for the lower and upper bounds of the approximated OP in the presence of the CEEs.

### C. Outage Probability in High SNR Region

In (21) and (22), suppose that $\zeta_l = \max_{1 \leq j \leq l} \{\frac{\gamma_{th_j}}{a_j - \gamma_{th_j} \sum_{i=j+1}^{L} a_i}\}$ is defined. Thus, $\mu_l = \frac{\zeta_l}{\gamma}$ and in the high SNR region $\alpha_2 \mu_l \to \sigma_{e,sr}^2 \zeta_l$ and $\alpha_1 \mu_l \to \sigma_{e,l}^2 \zeta_l$. Then, in the presence of the CEEs, *the lower and upper bounds of the approximated OP for the lth user in the high SNR region can be obtained, respectively as*

$$EF_{l,LB}^{app} = F_{\varphi_{sr}}(\sigma_{e,sr}^2 \zeta_l) + F_{\varphi_l}(\sigma_{e,l}^2 \zeta_l) F_{\varphi_{sr}}(\sigma_{e,sr}^2 \zeta_l) F_{\varphi_l}(\sigma_{e,l}^2 \zeta_l), \quad (23)$$

$$EF_{l,UB}^{app} = F_{\varphi_{sr}}(2\sigma_{e,sr}^2 \zeta_l) + F_{\varphi_l}(2\sigma_{e,l}^2 \zeta_l) F_{\varphi_{sr}}(2\sigma_{e,sr}^2 \zeta_l) F_{\varphi_l}(2\sigma_{e,l}^2 \zeta_l). \quad (24)$$

From (23) and (24), it is observed that EF bounds do not depend on the $\gamma$. Therefore, the OP reaches a fixed value in the high SNR region, which implies that the presence of CEE eliminates the diversity order of the system.

**Proposition 2:** In the absence of the CEEs, *the asymptotic OP for the lth user can be obtained in the high SNR region as follows:*

$$OP_{l,asymp}^{app} \approx \chi \mu_l^{\tau}, \quad (25)$$

where *for the lth user, the diversity order $DO_l$ and array gain $AG_l$ can be stated, respectively as*

$$DO_l = \tau = \min(\tau_1, \tau_2), \quad (26)$$

$$AG_l = (\chi)^{1/\tau}. \quad (27)$$

In (26), $\tau_1 = m_1 N_{t1} N_{r1}$, $\tau_2 = l m_2 N_{t2} N_{r2}$. By using that $\Omega = \Omega_1 = \eta \Omega_2$, $\eta$ is a positive constant, $\chi$ in (27) is given by

$$\chi = \begin{cases} \frac{(m_1/\Omega)^{\tau}}{\Gamma(m_1 N_{t1}+1)^{N_{r1}}} + \frac{L_l (m_2 \eta/\Omega)^{\tau}}{\Gamma(m_2 N_{t2}+1)^l}, & \tau_1 = \tau_2 = \tau \\ \frac{L_l (m_2 \eta/\Omega)^{\tau_2}}{\Gamma(m_2 N_{t2}+1)^l}, & \tau_1 > \tau_2 \\ \frac{(m_1/\Omega)^{\tau_1}}{\Gamma(m_1 N_{t1}+1)^{N_{r1}}}, & \tau_1 < \tau_2 \end{cases} \quad (28)$$

*Proof of Proposition 2:* In order to derive (25), we use high SNR approximation technique proposed in [39], i.e., $\gamma \to \infty$,

then $\mu_l \to 0$. In addition, by the property of incomplete Gamma function $\Psi(x, y \to 0) \approx y^x/x$ [5, eq. (45.9.1)], the asymptotic expressions $F_{\varphi_{sr}}^\infty$ and $F_{\varphi_l}^\infty$ can be stated, respectively as follows:

$$F_{\varphi_{sr}}^\infty(\mu_l) = \frac{(\frac{m_1\mu_l}{\Omega})^{m_1 N_{t1} N_{r1}}}{\Gamma(m_1 N_{t1}+1)^{N_{r1}}} = \frac{(m_1/\Omega)^{\tau_1}}{\Gamma(m_1 N_{t1}+1)^{N_{r1}}}\mu_l^{\tau_1}, \quad (29)$$

$$F_{\varphi_l}^\infty(\mu_l) = \frac{Q_l}{l}\frac{(\frac{m_2\eta\mu_l}{\Omega})^{lm_2 N_{t2} N_{r2}}}{\Gamma(m_2 N_{t2}+1)^l} = \frac{L_l\ (m_2\eta/\Omega)^{\tau_2}}{\Gamma(m_2 N_{t2}+1)^l}\mu_l^{\tau_2}, \quad (30)$$

By reforming $OP_l^{app}$ in (18) and using $\frac{ab}{a+b} \le \min(a,b)$ [39], the asymptotic OP for the CSI based gain AF relay can be given by

$$OP_{l,asymp}^{app}(\mu_l) = P_r(\min(\varphi_{sr},\varphi_l) < \mu_l) \simeq F_{\varphi_{sr}}^\infty(\mu_l) + F_{\varphi_l}^\infty(\mu_l). \quad (31)$$

By substituting (29) and (30) into (31), then the asymptotic OP of the $l$th user for the CSI based gain AF relay can be obtained as in (25).maj/MRC, the TAS-maj/MRC provides better performance than $A^3$/MRC and AIA/MRC schemes for two users out of three users.

## IV. NUMERICAL RESULTS

By this section, performance of cooperative NOMA network with MRT/RAS is presented by numerical results over Nakagami-$m$ fading channels. Consider $L = 3$ with power coefficients allocated as $a_1 = 3/6$, $a_2 = 2/6$ and $a_3 = 1/6$, the corresponding threshold SINR values of the three users can be considered as $\gamma_{th_1} = 0.9$, $\gamma_{th_2} = 1.5$ and $\gamma_{th_3} = 2$. Also, SNR threshold value of conventional OMA $\gamma_{th}$, which satisfies $\frac{1}{2}\sum_{i=1}^{L}\log_2(1+\gamma_{th_i}) = \frac{1}{2}\log_2(1+\gamma_{th})$, is used to have a comparison between the performance of conventional OMA and NOMA, and then the threshold value of conventional OMA is $\gamma_{th} = 13.25$. The path loss is assumed to be $\alpha = 4$. We assume $\epsilon_{sr} = \epsilon_l = \epsilon = 0.005$ in Fig. 2, 3 and 6 and $d_1 = d_2 = 0.5$ in all figures except Fig. 6.

Performance of the OP versus SNR is shown in Fig. 2. In the presence of the CEE values, the outage performance is plotted for various antenna configurations with equal Nakagami parameters for both hops, i.e., $(m_1, m_2) = (1,1)$, special case of Rayleigh fading. For exact OP simulations, (10) is used. And for the sake of clarity, the corresponding OP bounds of the $U_1$, i.e., the weakest user, are presented in Fig. 2-(b). As seen in Fig. 2-(a), the approximated analytical results are validated by the simulations and quite near to the exact ones. Moreover, the OP improves as the number of antennas increases. Although only the MRT improves the performance as compared to the single antenna case, the MRT/RAS provides enhancement significantly. Also, it is noticed that the outage performance for the users is different due to the interferences between the users. The OP $_1$ is the lowest one because it suffers from the interferences of the other users ($U_1$ and $U_2$) while the OP $_2$ is a little bit better than that of $U_1$ because it only suffers from the interference of $U_3$. On the other hand, the OP $_3$ achieves the best SNR as it does not suffer from any interferences of the other users. In Fig. 2-(b), the OP reaches an EF due to CEEs when the system SNR increases, which means zero diversity order.

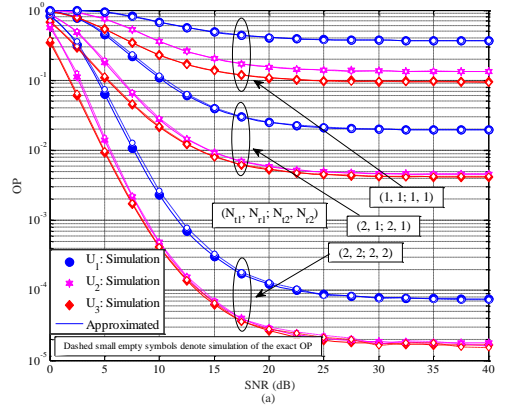

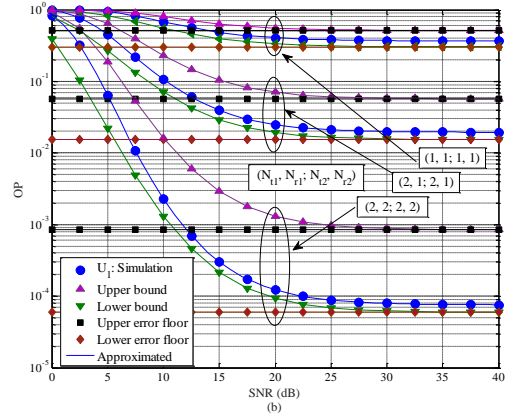

Figure 2: The OP of MRT/RAS-NOMA versus SNR for CSI based AF relay in case $(m_1, m_2) = (1, 1)$ and different antenna configurations.

Fig. 3 depicts the performance of the OP versus SNR for special case 1/RAS, i.e., $(N_{t1}, N_{r1}; N_{t2}, N_{r2}) = (1,2; 1,2)$, over different Nakagami parameters. It is noticed that the outage performance enhances as the channel conditions of the hops (values of Nakagami parameters) improve.

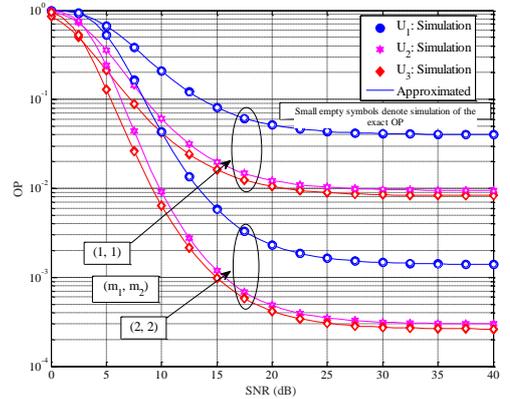

Figure 3: The OP of MRT/RAS-NOMA versus SNR for CSI based AF relay in case $(N_{t1}, N_{r1}; N_{t2}, N_{r2}) = (1, 2; 1, 2)$ and different Nakagami parameters.

Fig. 4 and Fig. 5 show effect of relative CEE on the outage performance. As observed in Fig. 4, the outage performance gets worse as the CEE increases. As seen in Fig. 5, the MRT/RAS scheme reaches an EF at an SNR of approximately 35 dB while the single antenna case does at around 25 dB. In addition, the EF values of around $10^{-1}$ and $10^{-7}$ occur for the single antenna case and MRT/RAS scheme, respectively. Thus, the MRT/RAS scheme is much more robust to CEEs than the single antenna case.

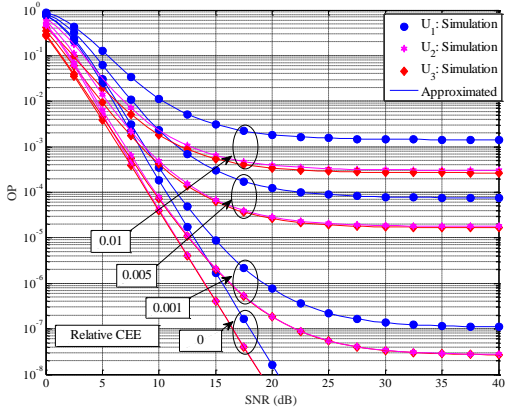

Figure 4: The OP of MRT/RAS-NOMA versus SNR for CSI based AF relay in case $(N_{t1}, N_{r1}; N_{t2}, N_{r2}) = (2, 2; 2, 2)$, $(m_1, m_2) = (1, 1)$ and different relative CEEs.

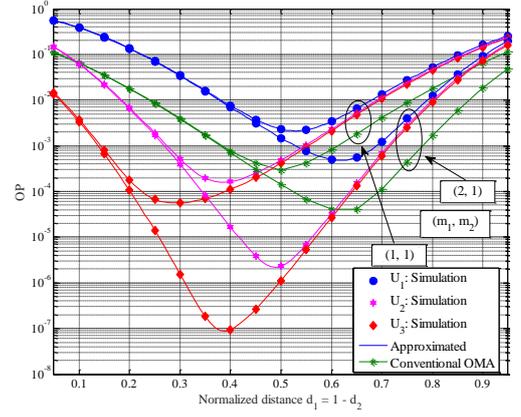

Figure 6: The OP of MRT/RAS-NOMA versus normalized distance $d_1$ for CSI based AF relay in case $(N_{t1}, N_{r1}; N_{t2}, N_{r2}) = (2, 2; 2, 2)$ and different Nakagami parameters.

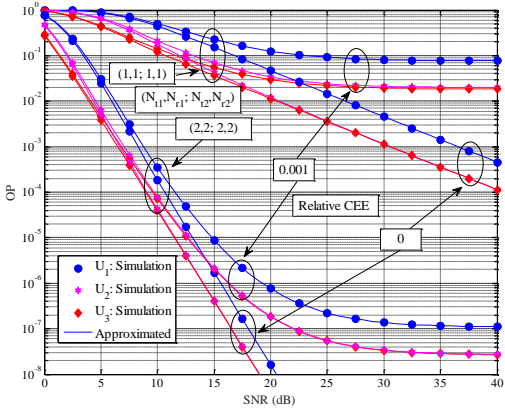

Figure 5: The OP of MRT/RAS-NOMA versus SNR for CSI based AF relay in case $(m_1, m_2) = (1, 1)$ and different antenna configurations and relative CEEs.

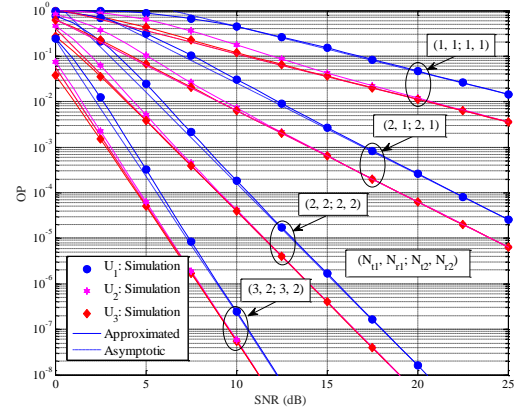

Figure 7: The OP of MRT/RAS-NOMA versus SNR for CSI based AF relay in case $(m_1, m_2) = (1, 1)$ and different antenna configurations.

Fig. 6 illustrates the OP performance versus the normalized distance between the BS and relay, $d_1$. The outage performance is investigated for $(N_{t1}, N_{r1}; N_{t2}, N_{r2}) = (2,2; 2,2)$ and different Nakagami parameters with SNR = 10 dB. As seen in Fig. 6, the optimum location of the relay is different and determined by three situations. In the first one, the relay has to be closer to the BS as the minimum OP is obtained at $d_1 < d_2$ when $m_1 N_{t1} N_{r1} < l m_2 N_{t2} N_{r2}$, such as the $U_2$ and $U_3$ in case $(m_1, m_2) = (1,1)$ and $U_3$ in case $(m_1, m_2) = (2,1)$. In the second situation, the relay has to be closer to the mobile users since the minimum OP is achieved at $d_1 > d_2$ when $m_1 N_{t1} N_{r1} > l m_2 N_{t2} N_{r2}$, such as the $U_1$ in case $(m_1, m_2) = (2,1)$. However, in the last one, the relay should be approximately in the center away from BS and mobile users, since the minimum OP is obtained at $d_1 \approx d_2$ in general when $m_1 N_{t1} N_{r1} = l m_2 N_{t2} N_{r2}$, such as $U_1$ in case $(m_1, m_2) = (1,1)$ and $U_2$ in case $(m_1, m_2) = (2,1)$. Fig. 6 also demonstrates the superiority of the NOMA with MRT/RAS over the conventional OMA with MRT/RAS when $d_1 < d_2$, and hence, the relay location for the NOMA with MRT/RAS prefer to be near the BS. In addition, although the conventional OMA has better performance than the proposed NOMA for $d_1 > d_2$, the NOMA has better spectral efficiency and user fairness than the conventional OMA.

The OP performance in the absence of the CEEs is plotted in Fig. 7 for the various antenna configurations with equal Nakagami parameters for both hops, i.e., $(m_1, m_2) = (1,1)$. The approximated analytical results are verified by the simulations. As noticed in Fig. 7, the outage performance improves considerably with the MRT/RAS scheme as compared to the single antenna case. In addition, the OP $_2$ and OP $_3$ behaviors at high SNR values become almost the same. Besides, the asymptotic curves are compatible with the exact ones and the diversity order expressed by (26) is validated.

## V. CONCLUSION

In this paper, the outage behavior of the MRT/RAS in the downlink NOMA network with AF relay has been analyzed over Nakagami-$m$ fading channels in the presence and absence of the CEEs. The approximated closed-form expression of the OP is derived and it is shown that it is too close to the exact one. Also, the lower and upper bounds of the OP are achieved. In the high SNR region, the OP reaches the EF due to the presence of the CEEs, and hence the diversity order fades while in the absence of the CEEs, better diversity order and array gain are achieved and is limited by the hop with smaller diversity order. Moreover, optimal relay location is different for mobile users and basically depends on the number of antennas and Nakagami parameters of the hops. Furthermore, the outage performance of the network becomes remarkable as the channel conditions improve or number of antennas increases. The MRT/RAS scheme provides significant performance and much more robustness to CEEs as compared to the single antenna case. Finally, the proposed NOMA with MRT/RAS assures better performance than the conventional OMA with MRT/RAS.